\documentclass[prd,aps,showpacs,showkeys,preprint,floats,eqsecnum,amsmath]{revtex4}
\usepackage{epsfig}

\begin{document}

\title{Forces and momenta caused by electromagnetic waves in magnetoelectric media}

\author{Yuri N.~Obukhov}
\email{yo@htp.uni-koeln.de}
\affiliation{Institute for Theoretical Physics, University of Cologne,
50923 K\"oln, Germany}
\affiliation{Department of Theoretical Physics,
Moscow State University, 117234 Moscow, Russia}

\author{Friedrich W. Hehl}
\email{hehl@thp.uni-koeln.de}
\affiliation{Institute for Theoretical Physics, University of Cologne,
50923 K\"oln, Germany}
\affiliation{Dept.\ of Phys.\ Astron., 
University of Missouri-Columbia, Columbia, MO 65211, USA}

\begin{abstract}
We analyse the propagation of electromagnetic waves in magnetoelectric 
media. Recently, Feigel has predicted that such a medium may ``extract 
momentum from vacuum" in the sense that the total momentum of the virtual
waves (vacuum fluctuations of the electromagnetic field) is nontrivial. 
Our aim is to check the feasibility of this effect. The crucial point in 
our study is an assumption of the finite size of the magnetoelectric sample, 
which allows us to reduce the calculation of the momenta and forces of the 
electromagnetic waves acting on the sample to the vacuum region outside of 
the medium. In this framework, we demonstrate that, in contrast to Feigel, 
the total force caused by the virtual is zero, with an appropriate count of 
the modes that should be taken into account in this effect. Furthermore, we 
find that the two irreducible parts of the magnetoelectric matrix behave 
differently in the possible Feigel effect. Going beyond the original scheme 
of the virtual electromagnetic waves, we propose an experimental scheme which 
is suitable for the measurement of the magnetoelectric susceptibilities of the 
medium with the help of {\it real} electromagnetic waves. 
\end{abstract}
\maketitle

%%%%%%%%%%%%%%%%%%%%%%%%%%%%%%%%%%%%%%%%%%%%%%%%%%%%%%%%%%%
\section{Introduction}
%%%%%%%%%%%%%%%%%%%%%%%%%%%%%%%%%%%%%%%%%%%%%%%%%%%%%%%%%%%

Phenomenological macroscopic electrodynamics is based on the well known
experimental observations that an external electric field can induce a 
polarization of a medium, whereas an external magnetic field can induce a
magnetization of matter. As a result, the electric and magnetic excitations
$({\cal D}, {\cal H})$ (comprising a 2-form $H$) are functions of the 
electric and magnetic field strengths $(E, B)$ (collected in a 2-form $F$) 
and of the permittivity $\varepsilon$ and the permeability $\mu$ of the medium. 
In the simplest case of an isotropic medium at rest, the constitutive relations 
read ${\cal D} = \varepsilon\varepsilon_0\,E$ and ${\cal H} = (\mu\mu_0)^{-1}B$. 
Here $\varepsilon_0$ and $\mu_0$ are the electric and magnetic constants 
(permittivity and permeability of the vacuum). However, in the 1960s it was 
theoretically predicted \cite{Dz1} and experimentally confirmed 
\cite{Astrov0,Astrov} that certain media become electrically polarized when 
placed into a magnetic field or are magnetized when put into an electric field. 
Since then, such a magnetoelectric effect was observed for many substances and 
studied in great detail both theoretically and experimentally, see for the reviews
\cite{ODell,Fiebig}, for example. The constitutive relation of a medium at rest 
is then modified to ${\cal D} = \varepsilon\varepsilon_0\,E + \beta\cdot B$ 
and ${\cal H} = (\mu\mu_0)^{-1}B - {\beta}^{\rm T}\cdot E$, where the traceless
$3\times 3$ matrix $\beta$ describes the magnetoelectric properties ($^{\rm T}$ 
denotes the transposed matrix). The magnetoelectric effect can be observed
in a certain class of media, but it also can be induced in media which
are put into external electric and magnetic fields. 

Recently, attention to magnetoelectric media was attracted in connection
with an interesting new effect predicted by Feigel \cite{feigel}. He noticed
that the propagation of electromagnetic waves in a magnetoelectric medium is 
essentially asymmetric in the sense that the waves moving in opposite directions 
carry different momenta. Then, calculating the total momentum of the virtual 
waves (or ``vacuum fluctuations" of the electromagnetic field) which are present 
in the sample, he concluded that this quantity is nontrivial. In this sense, 
the Feigel effect predicts the extraction of momentum from vacuum. This 
possibility was discussed in \cite{comm1,reply1,comm2,reply2,Tiggelen,Birkeland}.
In particular, in \cite{Birkeland} a certain similarity of the Feigel effect 
with the Casimir effect was noticed, and the computation of the total 
momentum transfered by the virtual waves was attacked by means of the Green's 
function method. In this paper, we analyse the Feigel effect in a sample of 
{\it finite} size which was somehow neglected in the previous works. 

It seems clear that the discussion of the Feigel effect is related to the
definition of the energy and momentum of the electromagnetic field in a 
medium, and moreover, in a moving medium (since a nontrivial velocity of
a sample is predicted \cite{feigel}). However, as it is well known, the issue 
of the energy and the momentum of the electromagnetic field (and of waves, 
in particular) in dielectric and magnetic media has a long and controversial 
history. The discussion of this issue began with the investigations of Minkowski 
\cite{minkowski}, Abraham \cite{abraham}, and Einstein and Laub \cite{laub}. 
The problem is reviewed in \cite{robin,Brevik,skob,ginz}, and
most recently, in \cite{newrmp}. However, till now the problem was not
settled neither theoretically, nor experimentally (see a discussion
in \cite{newrmp,tiersten,Dereli1,Dereli2}, e.g.). In the recent paper
\cite{emt} we have proposed a consistent definition of the electromagnetic
energy-momentum in media, and furthermore, a variational approach was developed
in \cite{emtme} for the moving magnetoelectric medium specifically for the 
study of the Feigel effect. An interesting technical observation is then
that the magnetoelectric matrix $\beta$ induces a term in the energy-momentum 
tensor which describes an additional flux of energy and momentum. 
Such an additional term is present even in the medium at rest, and it 
vanishes when the magnetoelectric properties are absent. This fact lends
some support to the feasibility of the Feigel effect, at least on a
qualitative level. 

Here, however, we will analyse the possibility of the Feigel effect 
avoiding the problem of the electromagnetic energy-momentum in the medium.
The crucial point is the assumption of a finite size of the magnetoelectric
sample. Then we notice that the virtual waves, which are excited in the medium,
are not confined to the sample. Since the boundaries are not impenetrable, 
the vacuum fluctuations exist everywhere, inside the medium as well as 
outside of it. Since the virtual electromagnetic waves in these regions
of space are related by the jump conditions across the boundaries, we can
eventually replace the evaluation of the electromagnetic momentum and 
forces inside the medium by the computation of these quantities in 
free space. In vacuum the energy-momentum is uniquely defined, and this
simplifies the analysis of the possible Feigel effect to a considerable
extent. 

The structure of the paper is as follows. Sec.~\ref{BM} presents some general
material, introduces the basic notions and fixes the notation. In Sec.~\ref{CR} 
we analyse the constitutive relation of the magnetoelectric medium. In particular, 
we provide an irreducible decomposition of the magnetoelectric matrix, which
proves to be convenient for the subsequent theoretical analysis. Sec.~\ref{Wave1}
demonstrates a general feature of the wave propagation in magnetoelectric
media, namely the birefringence. We demonstrate that for a medium 
characterized by the second irreducible piece, the generic Fresnel covector
surface factorizes into a product of two light cones. We find the 
two corresponding optical metrics. In Sec.~\ref{Wave2} the propagation of
plane electromagnetic waves is studied in a magnetoelectric medium of the 
Feigel type. From the jump conditions on the boundaries between the vacuum
regions and the medium, we derive relations between the transmitted and the
reflected waves. These relations are then used in Sec.~\ref{energydensity} 
for the calculation of the electromagnetic energy, the momentum and the force 
acting on the magnetoelectric sample.

%%%%%%%%%%%%%%%%%%%%%%%%%%%%%%%%%%%%%%%%%%%%%%%%%%%%%%%%%%%
\section{Some background material}\label{BM}
%%%%%%%%%%%%%%%%%%%%%%%%%%%%%%%%%%%%%%%%%%%%%%%%%%%%%%%%%%%

If the 4-dimensional (4D) electromagnetic {\it excitation} 2-form is denoted
by \footnote{Latin indices are coordinate indices. In 4D, they run over
$i,j,...=0,1,2,3$ and in 3D over $a,b,...=1,2,3$. The totally
antisymmetric Levi-Civita symbol in 4D is denoted by
$\hat{\epsilon}_{ijkl}=0,1,-1$ and in 3D by $\hat{\epsilon}_{abc}=0,1,-1$.}
\begin{equation}
   H =\!\! -\,{\cal H}\wedge d\sigma + {\cal D}=\frac 12
   H_{ij}dx^i\wedge dx^j
  \label{excit}
\end{equation}
and the 4D electromagnetic {\it field strength} 2-form by
\begin{equation}
  F =\!\! \quad E\wedge d\sigma
   + B=\frac 12 F_{ij}dx^i\wedge dx^j\,, \label{field}
\end{equation}\label{Max}
then the Maxwell equations read
\begin{equation}\label{Maxx}
dH=J\,,\qquad dF=0\,.
\end{equation}
Here $\sigma$ is a time parameter and ${\cal J}=-j\wedge d\sigma+\rho$
the 4D electric current. The 3D magnetic excitation ${\cal H}={\cal
   H}_adx^a$ is a 1-form, the electric excitation ${\cal D}=\frac 12 {\cal
   D}_{ab}dx^a\wedge dx^b$ a 2-form. Analogously, we have the 3D
electric field strength $E=E_a dx^a$ and the 3D magnetic field
strength $B=\frac 12 B_{ab}dx^a\wedge dx^b$. We also work with the
vector densities ${\cal D}^a=\frac 12 \epsilon^{abc}{\cal D}_{ab}$ and
$B^a=\frac 12 \epsilon^{abc}B_{ab}$.

We substitute (\ref{excit}) and  (\ref{field}) into (\ref{Maxx}), we
find the $1+3$ dimensional form of the Maxwell equations:
\begin{eqnarray}
   \underline{d}{\cal D}&=&\rho\,,\qquad \underline{d}{\cal H}-
\dot{\cal D}=j\,;\label{Maxi}\\
  \underline{d}B&=&0\,,\qquad  \underline{d}E+\dot{B}=0\,.
\end{eqnarray}
The 3D exterior derivative is denoted by $\underline{d}$ and the time
derivative by a dot. This is the premetric (i.e., metric independent)
framework of electrodynamics which summarizes the Maxwell equations
and their physical interpretation.

The properties of the medium under consideration are expressed by means
of the constitutive law. With the assumptions of locality and
linearity, we have
\begin{equation}\label{const}
   H_{ij}=\frac 12 \kappa_{ij}{}^{kl}F_{kl}=
\frac 14 \hat{\epsilon}_{ijmn}\chi^{mnkl}F_{kl}\,.
\end{equation}
It is convenient to introduce, besides the constitutive tensor density
$\kappa_{ij}{}^{kl}$, the tensor $\chi^{ijkl}$, since the latter is
used conventionally in electrodynamics, see \cite{Post}. Its
symmetries are $\chi^{ijkl}=-\chi^{jikl}=-\chi^{ijlk}$, i.e., it has
36 independent components.

The constitutive law can be put into different forms in order to
customize it for different applications. If we put it into a $6\times6$
form
\begin{equation}
   \left(\begin{array}{c} {\cal H}_a \\ {\cal D}^a\end{array}\right) =
   \left(\begin{array}{cc} {{\cal  C}}^{b}{}_a & {{\cal  B}}_{ba} \\
       {{\cal  A}}^{ba}& {{\cal  D}}_{b}{}^a \end{array}\right)
   \left(\begin{array}{c} -E_b\\ {B}^b\end{array}\right)\,,\label{CR'}
\end{equation}
then the $3\times 3$ matrices ${\cal A,B,C,D}$ can be related to the
4-dimensional constitutive tensor density $\chi^{ijkl}$ by
\begin{eqnarray}\label{A-matrix0}
   {\cal  A}^{ba}&=& \chi^{0a0b}\,,\quad
{\cal  B}_{ba}=\frac{1}{4}\,\hat\epsilon_{acd}\,
\hat\epsilon_{be\!f} \,\chi^{cdef}\,,\\
\label{C-matrix0}
{\cal  C}^a{}_b& =&\frac{1}{2}\,\hat\epsilon_{bcd}\,\chi^{cd0a}\,,\quad
{\cal  D}_a{}^b=\frac{1}{2}\,\hat\epsilon_{acd}
\,\chi^{0bcd}\,.
\end{eqnarray}
Then  $6\times 6$ form of $\chi^{ijkl}$ can be written as
\begin{equation}\label{kappachi}
\chi^{IK}= \left( \begin{array}{cc} {\cal B}_{ab}& {\cal D}_a{}^b \\
{\cal C}^a{}_b & {\cal A}^{ab} \end{array}\right)\,,
\end{equation}
with $I,K,...=01,02,03,23,31,12$.

We can decompose $\chi^{ijkl}$ irreducibly under the linear group. As
we have shown elsewhere \cite{HO02}, we find
\begin{equation}
   \underbrace{ \chi^{ijkl}}_{36} =
   \underbrace{{}^{(1)}\chi^{ijkl}}_{principal\> 20} +\underbrace{
     {}^{(2)}\chi^{ijkl}}_{skewon\> 15} + \underbrace{
     {}^{(3)}\chi^{ijkl}}_{axion\> 1}\label{chi-dec}\,.
\end{equation}
The irreducible pieces carry the additional symmetries
\begin{equation}\label{symm}
  {}^{(1)}\chi^{ijkl}= {}^{(1)}\chi^{klij}\,,\quad
  {}^{(2)}\chi^{ijkl}=- {}^{(2)}\chi^{klij}\,,\quad
  {}^{(3)}\chi^{ijkl}= {}^{(3)}\chi^{[ijkl]}\,.
\end{equation}
The {\it principal\/} part with its 20 independent components is the
only one discussed conventionally. The {\it skewon\/} part with its 15
components vanishes if one assumes the existence of a Lagrangian
4-form from which the constitutive law can be derived completely.
Finally, the {\it axion\/} piece with only 1 independent component is
totally antisymmetric:
\begin{equation}
  {}^{(3)}\chi^{ijkl}=\alpha\,\epsilon^{ijkl}\,.
\end{equation}
The $\alpha$ is a 4D pseudoscalar.

We take care of the irreducible decomposition and evaluate the
matrix elements of (\ref{CR'}):
\begin{eqnarray}\label{explicit'}
   {\cal H}_a\!&=\!&\left( \mu_{ab}^{-1} - \hat{\epsilon}_{abc}m^c
   \right) {B}^b +\left(- \beta^b{}_a + s_a{}^b - \delta_a^b
     s_c{}^c\right)E_b - \alpha\,E_a\,,\\ {\cal D}^a\!&=\!&\left(
     \varepsilon^{ab}\hspace{4pt} - \, \epsilon^{abc}\,n_c
   \right)E_b\,+\left(\hspace{9pt} \beta^a{}_b + s_b{}^a - \delta_b^a
     s_c{}^c\right) {B}^b + \alpha\,B^a \,.\label{explicit2}
\end{eqnarray}
We have $\varepsilon^{ab}=\varepsilon^{ba}$,
$\mu^{-1}_{ab}=\mu^{-1}_{ba}$, and $\beta^c{}_c=0$. Thus we have the
independent components of $\varepsilon^{ab}$ (6), $\mu^{-1}_{ab}$ (6),
$\beta^a{}_b$ (8), $m^c$ (3), $n_c$ (3), $s_a{}^b$ (9), and $\alpha$
(1). This adds up, as it is required, to 36.

In this paper, we assume that there exists a Lagrangian from which the
constitutive law can be derived. Thus, ${}^{(2)}\chi^{ijkl}=0$.
Moreover, we assume a vanishing axion part $\alpha=0$. Still, in
certain substances an axion part can be present, as we discussed
recently, see \cite{ax1,ax2}. After these ``amputations'', the
constitutive law to be investigated reads
\begin{eqnarray}\label{explicit''}
   {\cal H}_a\!&=\!&\mu_{ab}^{-1} {B}^b - \beta^b{}_a E_b\,,\\
   {\cal D}^a\!&=\!&
   \varepsilon^{ab} E_b\,\hspace{4pt}+ \beta^a{}_b {B}^b\,.
\end{eqnarray}
Recall that $\beta^c{}_c=0$.

%%%%%%%%%%%%%%%%%%%%%%%%%%%%%%%%%%%%%%%%%%%%%%%%%%%%%%%%%%%
\section{Constitutive relation for magnetoelectric media}\label{CR}
%%%%%%%%%%%%%%%%%%%%%%%%%%%%%%%%%%%%%%%%%%%%%%%%%%%%%%%%%%%

Throughout the paper we will use the exterior calculus which proves to
be very effective and convenient both in describing the general formalism
and in the specific computations. Here we put the constitutive relation
for the magnetoelectric media into a simple and transparent form by using 
the language of exterior calculus. 

As is worked out in \cite{HO02}, magnetoelectric properties of the medium 
are described by a tracefree $3\times 3$ matrix $\beta^a{}_b$ in the rest 
frame. The corresponding spacetime foliation is called the {\it laboratory} 
foliation, with the coordinate time variable $\sigma$ labeling the slices of 
this foliation. The spacetime metric ${\mathbf g}$ introduces a scalar 
product in the tangent space and defines the line element which reads 
($a,b,... = 1,2,3$)
\begin{equation}
ds^2 = N^2\,d\sigma^2 + g_{ab}\,\underline{dx}^a\,\underline{dx}^b 
= N^2\,d\sigma^2 - {}^{\hbox{$\scriptstyle{(3)}$}}g_{ab}
\,\underline{dx}^a\,\underline{dx}^b.\label{metF}
\end{equation} 
Here $N^2 = {\mathbf g}(n,n)$ is the length square of the foliation vector
field $n$, and $\underline{dx}^a = dx^a - n^a\,d\sigma$ is the transversal
3-covector basis, in accordance with the definitions above. The 3-metric
${}^{\hbox{$\scriptstyle{(3)}$}}g_{ab}$ is the positive definite Riemannian
metric on the spatial 3-dimensional slices corresponding to fixed values 
of the time $\sigma$. This metric defines a 3-dimensional Hodge duality
operator ${}^{\underline{\star}}$. Modern discussion of the classical 
electrodynamics, in particular, using the exterior calculus, can be found
in \cite{Post,Toupin,HO02,Lindell,Delph}.

Following Feigel, we do not consider the effects of gravity. Accordingly, we 
are in the Minkowski spacetime and a convenient choice of the laboratory 
foliation is $\sigma =t$ and $n^a = 0$ (hence $\underline{dx}^a = dx^a$). 
The spatial metric is Euclidean, ${}^{\hbox{$\scriptstyle{(3)}$}}g_{ab} 
= {\rm diag}(1,1,1)$, and $N =c$. 

We introduce the 1-form $\beta^a = \beta^a{}_b dx^b$. Then the constitutive
relation for the magnetoelectric medium, provided we assume isotropic 
permittivity and permeability, reads
\begin{eqnarray}
{\cal D} &=& \varepsilon\varepsilon_0\,{}^{\underline{\star}}E
- \beta^a\wedge e_a\rfloor B,\label{constM1}\\
{\cal H} &=& {\frac 1 {\mu\mu_0}}\,{}^{\underline{\star}}B 
- \beta^a\wedge e_a\rfloor E.\label{constM2}
\end{eqnarray}
We can decompose the tracefree $\beta^a$ into two irreducible parts 
(symmetric, and antisymmetric):
\begin{equation}\label{betadec}
\beta^a = {}^{(1)}\beta^a + {}^{(2)}\beta^a %+ {}^{(3)}\beta^a
= {}^{(1)}\beta^a + {}^{\underline{\star}}(\check{\beta}\wedge dx^a).
\end{equation}
The antisymmetric (pseudotrace) part is defined by ${}^{(2)}\beta^a : = 
{}^{\underline{\star}}(\check{\beta}\wedge dx^a)$ with a 1-form $\check{\beta} 
:= {\frac 13}{}^{\underline{\star}}(dx_a\wedge\beta^a)$. Obviously, 
$e_a\rfloor {}^{(2)}\beta^a = 0$.  Finally, the first 
irreducible part is trace- and pseudotrace-free, i.e., $e_a\rfloor {}^{(1)}
\beta^a = 0$ and $dx_a\wedge {}^{(1)}\beta^a = 0$. 

The first term in (\ref{betadec}) describes the tracefree symmetric part of 
the matrix $\beta^a{}_b$, whereas the second term is its antisymmetric part. 
When the first irreducible piece vanishes, the constitutive relation 
(\ref{constM1})-(\ref{constM2}) becomes much simpler:
\begin{eqnarray}
{\cal D} &=& \varepsilon\varepsilon_0\,{}^{\underline{\star}}E
+ {}^{\underline{\star}}B\wedge \check{\beta},\label{constMa1}\\
{\cal H} &=& {\frac 1 {\mu\mu_0}}\,{}^{\underline{\star}}B 
+ {}^{\underline{\star}}(E\wedge \check{\beta}).\label{constMa2}
\end{eqnarray}

In this paper, we will mainly analyse the general case 
(\ref{constM1})-(\ref{constM2}), with special attention to the medium
studied by Feigel. The latter is characterized by the magnetoelectric matrix
\begin{equation}\label{constF}
\beta^a{}_b = \left(\begin{array}{ccc} 0& 0& 0\\ 0& 0& \beta^2{}_3\\
0 & \beta^3{}_2 & 0\end{array}\right).
\end{equation}
Such magnetoelectric susceptibilities can be induced in an ordinary medium
(characterized by the permittivity $\varepsilon$ and permeability $\mu$)
under the action of external electric and magnetic fields applied along the 
second and third axes. In spite of the  simple form of (\ref{constF}), there 
are still two nontrivial irreducible parts. When $\beta^2{}_3 = \beta^3{}_2$, 
the pseudotrace $\check{\beta}$ vanishes, whereas for $\beta^2{}_3 = - 
\beta^3{}_2$ the first irreducible piece disappears. 

Our subsequent analysis reveals that different irreducible parts of
the magnetoelectric matrix are differently involved into the possible 
Feigel effect. 

The magnetoelectric matrix has the dimension $[\beta^a{}_b] = [\sqrt{
\varepsilon_0/\mu_0}]$. Accordingly, it is convenient to introduce a 
dimensionless object $\overline{\beta}^a{}_b := \beta^a{}_b/\lambda$ where
$\lambda = \sqrt{\varepsilon\varepsilon_0/\mu\mu_0}$. We will use the same
``overlined" notation also for the various exterior forms constructed from 
the magnetoelectric matrix.

%%%%%%%%%%%%%%%%%%%%%%%%%%%%%%%%%%%%%%%%%%%%%%%%%%%%%%%%%%%
\section{Birefringence in magnetoelectric media}\label{Wave1}
%%%%%%%%%%%%%%%%%%%%%%%%%%%%%%%%%%%%%%%%%%%%%%%%%%%%%%%%%%%

The Fresnel approach (geometric optics) to the wave propagation in media 
and in spacetime with the general linear constitutive law (\ref{const})
gives rise to the {\it extended covariant Fresnel 
equation} \cite{HO02} for the wave covector $q_i$:
\begin{equation} \label{Fresnel}  
{\cal G}^{ijkl}(\chi)\,q_i q_j q_k q_l = 0 \,.
\end{equation}
Here the fourth order Tamm-Rubilar (TR) tensor density of weight $+1$ is 
constructed from the constitutive tensor $\chi^{ijkl}$
as a cubic contraction with the Levi-Civita densities:
\begin{equation}\label{G4}  
  {\cal G}^{ijkl}(\chi):=\frac{1}{4!}\,\hat{\epsilon}_{mnpq}\,
  \hat{\epsilon}_{rstu}\, {\chi}^{mnr(i}\, {\chi}^{j|ps|k}\,
  {\chi}^{l)qtu }\,.
\end{equation}
Using the $(1+3)$-decomposed representation in terms of the $3\times 3$-matrices
(\ref{kappachi}), the independent components of the TR-tensor 
(\ref{G4}) read explicitly as follows:
\begin{eqnarray}  
 \label{ma0} M \!\!&:=&\!\! {\cal G}^{0000} = \det{\cal A} \,,\\
M^a \!\!&:=&\!\! 4\,{\cal G}^{000a} = -\hat{\epsilon}_{bcd}\left( {\cal A}^{ba}
\,{\cal A}^{ce}\,{\cal C}^d_{\ e} + {\cal A}^{ab}\,{\cal A}^{ec}
\,{\cal D}_e^{\ d}\right)\,,\label{ma1}\\
 M^{ab} \!\!&:=&\!\! 6\,{\cal G}^{00ab} = \frac{1}{2}\,{\cal A}^{(ab)}\left[
({\cal C}^d{}_d)^2 + ({\cal D}_c{}^c)^2 - ({\cal C}^c{}_d + {\cal D}_d{}^c)
({\cal C}^d{}_c +  {\cal D}_c{}^d)\right]\nonumber\\ 
&& +\,({\cal C}^d{}_c + {\cal D}_c{}^d)({\cal A}^{c(a}{\cal C}^{b)}{}_d 
+ {\cal D}_d{}^{(a}{\cal A}^{b)c}) - {\cal C}^d{}_d{\cal A}^{c(a}{\cal C}^{b)}
{}_c\nonumber\\ && -\,{\cal D}_c{}^{(a}{\cal A}^{b)c}{\cal D}_d{}^d -
  {\cal A}^{dc}{\cal C}^{(a}{}_c {\cal D}_d{}^{b)} + 
  \left({\cal A}^{(ab)}{\cal A}^{dc}- {\cal A}^{d(a}{\cal A}^{b)c}\right)
   \!{\cal B}_{dc},\label{ma2}\\
M^{abc} \!\!&:=&\!\! 4\,{\cal G}^{0abc} =
  \epsilon^{de(c|}\left[{\cal B}_{df}( {\cal A}^{ab)}\,{\cal D}_e^{\ f} 
 - {\cal D}_e^{\ a}{\cal A}^{b)f}\,) \right. \nonumber \\
&&\left.+\,{\cal B}_{fd}({\cal A}^{ab)}\,{\cal C}_{\ e}^f - {\cal A}^{f|a}
{\cal C}_{\ e}^{b)}) +{\cal C}^{a}_{\ f}
\,{\cal D}_e^{\ b)}\,{\cal D}_d^{\ f} + {\cal D}_f^{\ a}
\,{\cal C}^{b)}_{\ e}\,{\cal C}^{f}_{\ d} \right] \,,\label{ma3}\\
M^{abcd} \!\!&:=&\!\! {\cal G}^{abcd} = \epsilon^{ef(c}\epsilon^{|gh|d}
\,{\cal B}_{hf}\left[\frac{1}{2} \,{\cal A}^{ab)}\,{\cal B}_{ge} 
- {\cal C}^{a}_{\  e}\,{\cal D}_g^{\ b)}\right] \,.\label{ma4}
\end{eqnarray}
Then, in $(1+3)$-decomposed form, the extended Fresnel equation
(\ref{Fresnel}) reads
\begin{equation}
  q_0^4 M + q_0^3q_a\,M^a + q_0^2q_a q_b\,M^{ab} + q_0q_a q_b
  q_c\,M^{abc} + q_a q_b q_c q_d\,M^{abcd}=0\,.\label{decomp}
\end{equation} 

In Minkowski spacetime, for the general linear constitutive relation 
(\ref{constM1})-(\ref{constM2}) we have explicitly
\begin{equation}
{\cal A}^{ab} = -\,\varepsilon\varepsilon_0\,\delta^{ab},\quad
{\cal B}_{ab} = {\frac 1{\mu\mu_0}}\,\delta_{ab},\quad {\cal C}^a{}_b =
{\cal D}_b{}^a = \beta^a{}_b.
\end{equation}

In the {\it special} case (\ref{constMa1})-(\ref{constMa2}) when only the 
irreducible antisymmetric (pseudotrace) part of the magnetoelectric matrix 
is nontrivial, we analysis of the Fresnel equation (\ref{decomp}) reveals 
the clear birefringence effect. Indeed, using the constitutive relation 
(\ref{constMa1})-(\ref{constMa2}) in (\ref{ma0})-(\ref{ma4}), we find 
explicitly (with $n = \sqrt{\varepsilon\mu}$ as the usual index of refraction,
and $c = 1/\sqrt{\varepsilon_0\mu_0}$ the vacuum speed of light):
\begin{eqnarray}
M &=& -\,\left({\frac n c}\right)^3,\qquad M^a = 4\left({\frac n c}\right)^2
\overline{\beta}{}^a,\\
M^{ab} &=& {\frac n c}\left[\delta^{ab}(2 + \overline{\beta}{}^2) - 
5\overline{\beta}{}^a\overline{\beta}{}^b\right],\\
M^{abc} &=& -\,2\delta^{(ab}\overline{\beta}{}^{c)}\,(2 + \overline{\beta}{}^2)
+ 2\overline{\beta}{}^a\overline{\beta}{}^b\overline{\beta}{}^c,\\
M^{abcd} &=& {\frac c n}\left[- \delta^{(ab}\delta^{cd)}(1 + \overline{\beta}{}^2)
+ \delta^{(ab}\overline{\beta}{}^c\overline{\beta}{}^{d)}\right]. 
\end{eqnarray}
Here from the magnetoelectric 1-form $\check{\beta}$ we extract the 
dimensionless covector with the components $\overline{\beta}{}_a :=
e_a \rfloor\check{\beta}/\lambda$. Further, we denote $\overline{\beta}{}^a 
:= \delta^{ab}\overline{\beta}{}_b$ and $\overline{\beta}{}^2 :=
\overline{\beta}{}^a\overline{\beta}{}_a$. 

Substituting this into the Fresnel equation (\ref{Fresnel}), (\ref{decomp}),
we find {\it birefringence}: the quartic wave surface is 
factorized into the product of the two light cones,
\begin{equation}
{\cal G}^{ijkl}(\chi)\,q_i q_j q_k q_l = -\,\left(g_1^{ij}q_iq_j\right)
\,\left(g_2^{kl}q_kq_l\right) = 0.\label{2cones1}
\end{equation}
The two optical metrics \cite{gordon} depend explicitly on the magnetoelectric
properties according to
\begin{equation}
g_1^{ij} = \left(\begin{array}{c|c} {\frac {n^2}{c^2}} &  - {\frac nc}
\overline{\beta}{}^b \\ \hline - {\frac nc}\overline{\beta}{}^a & -\delta^{ab}
\end{array}\right)\label{opt1}
\end{equation}
and
\begin{equation}\label{opt2}
g_2^{ij} = \left(\begin{array}{c|c} {\frac {n^2}{c^2}} &  - {\frac nc}
\overline{\beta}{}^b \\ \hline - {\frac nc}\overline{\beta}{}^a & -\delta^{ab}(1 + 
\overline{\beta}{}^2) + \overline{\beta}{}^a\overline{\beta}{}^b\end{array}\right),
\end{equation}
respectively. It is interesting that the magnetoelectric covector manifests 
itself as an effective ``rotation" of the spacetime related to the 
off-diagonal components of the optical metric.

In the general linear case of the constitutive relation 
(\ref{constM1})-(\ref{constM2}),
the quartic Fresnel surface does not factorize into the product of the two 
light cones, in general. The birefringence effect is much more nontrivial 
in this case. We, however, will eventually specialize to the medium 
(\ref{constF}) studied by Feigel, and in this case we can analyse the 
propagation of the planes waves to the end. 

%%%%%%%%%%%%%%%%%%%%%%%%%%%%%%%%%%%%%%%%%%%%%%%%%%%%%%%%%%%
\section{Plane waves in a magnetoelectric medium}\label{Wave2} 
%%%%%%%%%%%%%%%%%%%%%%%%%%%%%%%%%%%%%%%%%%%%%%%%%%%%%%%%%%%

A general discussion of the propagation of electromagnetic waves in
media with a local and linear constitutive relation can be found in
\cite{ODell,Serdyukov,LindellSihvola,HO02}. Various special cases of the 
constitutive tensor were analysed, including also the magnetoelectric cases.
Here we will confine our attention to the specific form of the magnetoelectric
matrix (\ref{constF}). 

%%%%%%%%%%%%%%%%%%%%%%%%%%%%%%%%%%%%%%%%%%%%%%%%%%%%%%%%%%%%%%%%%%%%%%%
\begin{figure}
\epsfxsize=\hsize 
\epsfbox{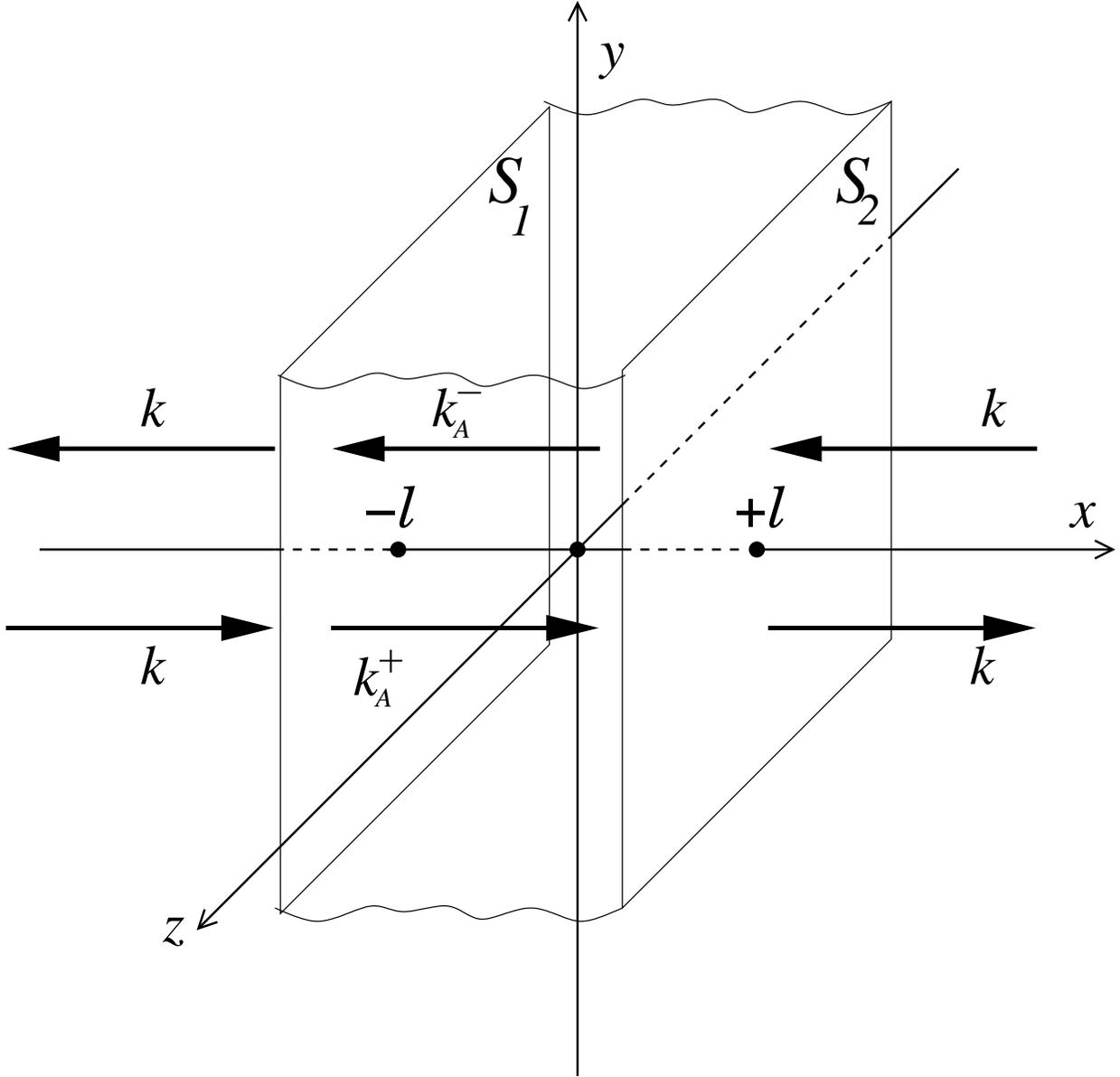}
\caption[An infinitely extended (in $y$ and $z$ directions) slab of a
magnetoelectric medium is located between the plane boundaries $S_1$ and $S_2$.
We study the propagation of electromagnetic waves in $x$-direction inside
and outside the magnetoelectric medium.] {\label{plane}{An infinitely extended 
(in $y$ and $z$ directions) slab of a
magnetoelectric medium is located between the plane boundaries $S_1$ and $S_2$.
We study the propagation of electromagnetic waves in $x$-direction inside
and outside the magnetoelectric medium.}}
\end{figure}
%%%%%%%%%%%%%%%%%%%%%%%%%%%%%%%%%%%%%%%%%%%%%%%%%%%%%%%%%%%%%%%%%%%%%%%

In order to clarify the possible Feigel effect, we need to analyse not
only the waves inside the sample (as was done originally in \cite{feigel})
but also the waves in the outside vacuum space. The appropriate qualitative 
picture is as follows (see Fig.~\ref{plane}): Let us put the magnetoelectric 
matter between the two parallel planes $S_1 = \{x = -\ell\}$ and $S_2 = \{x = 
+\ell\}$. Feigel considers the case when the magnetoelectric properties are 
induced by {\it external} electric and magnetic fields. When these external 
crossed electric and magnetic fields are applied parallelly to the boundaries 
$S_1$ and $S_2$, the magnetoelectric matrix will have the form (\ref{constF}).
The idea of Feigel is that the virtual electromagnetic waves (vacuum 
fluctuations of the electromagnetic field) propagating
in such a magnetoelectric medium could produce a nontrivial momentum 
along the distinguished axis $x$. However, the presence of similar
virtual waves outside of the medium was not taken into account (despite
the fact that the magnetoelectric sample was assumed to have finite size).
Here we reanalyse carefully the picture, taking into account all the 
virtual electromagnetic waves inside and outside of the medium alike.

We begin by noticing that outside of the matter, we have the ``bath" of the 
virtual photons some of which will penetrate the interior of the sample, 
reflecting and refracting at its boundaries. Obviously, the largest 
contribution to the possible effect should come from the electromagnetic 
waves which travel along the $x$-axis, i.e., with the wave vectors normal 
to the boundaries. Clearly, for each right-moving wave falling on the left 
boundary $S_1$, there exists an equal but opposite left-moving wave falling 
on the right boundary $S_2$. The contributions of these {\it ingoing} waves to 
the momentum density of the electromagnetic field are equal with opposite 
sign, thus providing a {\it balance} of the light pressures in the left and 
in the right vacuum regions. However, we have to find the contributions of 
the {\it outgoing} waves. If they turn out to be different in the left and 
in the right vacuum regions, this would seemingly yield a violation of the 
momentum balance and would encompass a nontrivial Feigel effect. 

\begin{table}
\caption{\label{table1}Amplitudes in the different regions and
   direction of the corresponding wave. The polarization index 1 and 2
   describes a wave with the electric field in y- and in z-direction,
   respectively} 
\begin{ruledtabular}
\begin{tabular}{|c|l|l|}\hline
  region 1&region 3& region 2\\ \hline\hline
  $a_1,a_2$&$p_1,p_2$&$c_1,c_2$\\
$\longrightarrow$&$\longrightarrow$&$\longrightarrow$\\ \hline
  $b_1,b_2$&$q_1,q_2$&$d_1,d_2$\\
$\longleftarrow$&$\longleftarrow$&$\longleftarrow$\\ \hline
\end{tabular}
\end{ruledtabular}
\end{table}

There are three regions (see Fig.~\ref{plane} and Table~\ref{table1}): 
1) the left vacuum space (for $x < -\ell$), 2) the right vacuum space (for 
$x > \ell$), and 3) the interior region filled with magnetoelectric matter 
(for $-\ell < x < \ell$). The configurations of the electromagnetic field 
in these three domains read, respectively, as follows, where 
we use the complex representation to simplify the formulas:

1) In the first region ($x < -\ell$): 
\begin{eqnarray}\label{E1}
E &=& e^{-i\omega t}\left[\left(a_1e^{ikx} + b_1e^{-ikx}\right)dy
+ \left(a_2e^{ikx} + b_2e^{-ikx}\right)dz\right],\\ \label{B1}
B &=& {\frac k\omega}\,e^{-i\omega t}\,dx\wedge\left[\left(a_1e^{ikx} 
- b_1e^{-ikx}\right)dy + \left(a_2e^{ikx} - b_2e^{-ikx}\right)dz\right],\\
{\cal D} &=& \varepsilon_0\,e^{-i\omega t}\,dx\wedge\left[-\left(a_1e^{ikx} + 
b_1e^{-ikx}\right)dz + \left(a_2e^{ikx} + b_2e^{-ikx}\right)dy\right],\label{D1}\\
{\cal H} &=& {\frac k{\mu_0\omega}}\,e^{-i\omega t}\left[\left(a_1e^{ikx} 
- b_1e^{-ikx}\right)dz - \left(a_2e^{ikx} - b_2e^{-ikx}\right)dy\right].\label{H1}
\end{eqnarray}
Here the complex amplitudes $a_1$ and $a_2$ describe the right-moving waves 
with the two different polarizations (the subscript $_1$ refers to the first
polarization with the electric field along the $y$-axis, whereas the subscript
$_2$ denotes a second independent polarization with the electric field along 
the $z$-axis). The complex amplitudes $b_1$ and $b_2$ describe the corresponding 
left-moving waves. With $k = \omega/c$ one can straightforwardly check that 
this configuration is a solution of the Maxwell equations.

2) Similarly, in the second region ($x > \ell$):
\begin{eqnarray}\label{E2}
E &=& e^{-i\omega t}\left[\left(c_1e^{ikx} + d_1e^{-ikx}\right)dy
+ \left(c_2e^{ikx} + d_2e^{-ikx}\right)dz\right],\\ \label{B2}
B &=& {\frac k\omega}\,e^{-i\omega t}\,dx\wedge\left[\left(c_1e^{ikx} 
- d_1e^{-ikx}\right)dy + \left(c_2e^{ikx} - d_2e^{-ikx}\right)dz\right],\\
{\cal D} &=& \varepsilon_0\,e^{-i\omega t}\,dx\wedge\left[-\left(c_1e^{ikx} + 
d_1e^{-ikx}\right)dz + \left(c_2e^{ikx} + d_2e^{-ikx}\right)dy\right],\label{D2}\\
{\cal H} &=& {\frac k{\mu_0\omega}}\,e^{-i\omega t}\left[\left(c_1e^{ikx} 
- d_1e^{-ikx}\right)dz - \left(c_2e^{ikx} - d_2e^{-ikx}\right)dy\right].\label{H2}
\end{eqnarray}
Now the amplitudes $c_1, c_2$ and $d_1, d_2$ describe the right- and the
left-moving waves, respectively, with the two polarizations, and again 
$k = \omega/c$.

3) In the third (interior) region (with $-\ell < x < \ell$), the field 
configurations look somewhat more nontrivial:
\begin{eqnarray}\label{E3}
E &=& e^{-i\omega t}\left[\left(p_1e^{ik_1^+x} + q_1e^{-ik_1^-x}\right)dy
+ \left(p_2e^{ik_2^+x} + q_2e^{-ik_2^-x}\right)dz\right],\\ \label{B3}
B &=& {\frac {e^{-i\omega t}}\omega}\,dx\wedge\left[\left(k_1^+p_1e^{ik_1^+x} 
- k_1^-q_1e^{-ik_1^-x}\right)dy + \left(k_2^+p_2e^{ik_2^+x} - k_2^-q_2
e^{-ik_2^-x}\right)dz\right],\\
{\cal D} &=& e^{-i\omega t}\,dx\wedge\Big[-\left(\varepsilon\varepsilon_0 
+ {\frac {k_1^+}\omega}\beta^2{}_3\right)p_1e^{ik_1^+x}dz + \left(\varepsilon
\varepsilon_0 - {\frac {k_2^+}\omega}\beta^3{}_2\right)p_2e^{ik_2^+x}dy\nonumber\\
&& - \left(\varepsilon\varepsilon_0 - {\frac {k_1^-}\omega}\beta^2{}_3\right)
q_1e^{-ik_1^-x}dz + \left(\varepsilon\varepsilon_0 + {\frac {k_2^-}\omega}
\beta^3{}_2\right)p_2e^{-ik_2^-x}dy\Big],\label{D3}\\
{\cal H} &=& e^{-i\omega t}\Big[\left({\frac {k_1^+}{\mu\mu_0\omega}} - 
\beta^2{}_3\right)p_1e^{ik_1^+x}dz - \left({\frac {k_2^+}{\mu\mu_0\omega}}
+ \beta^3{}_2\right)p_2e^{ik_2^+x}dy \nonumber\\
&& - \left({\frac {k_1^-}{\mu\mu_0\omega}} + \beta^2{}_3\right)q_1e^{-ik_1^-x}dz 
+ \left({\frac {k_2^-}{\mu\mu_0\omega}} - \beta^3{}_2\right)q_2e^{-ik_2^-x}dy
\Big].\label{H3}
\end{eqnarray}
The matter is characterized by the permittivity $\varepsilon$, the 
permeability $\mu$, and the magnetoelectric matrix $\beta^a{}_b$ 
(the latter is given by (\ref{constF})). The right- and the left-movers are 
now described by complex amplitudes $p_1, p_2$ and $q_1, q_2$, respectively. 
The birefringence, however, is manifest in the inequality of $k_1^\pm \neq 
k_2^\pm$: the waves with different polarization have different propagation 
vectors. Explicitly, we find
\begin{eqnarray}
k_1^{\pm} &=& {\frac {n\omega}{c}}\left(\sqrt{(\overline{\beta}{}^2{}_3)^2 + 1} 
\pm\overline{\beta}{}^2{}_3\right),\label{k1}\\
k_2^{\pm} &=& {\frac {n\omega}{c}}\left(\sqrt{(\overline{\beta}{}^3{}_2)^2 + 1} 
\mp\overline{\beta}{}^3{}_2\right),\label{k2}\end{eqnarray}
As before, we use here the dimensionless magnetoelectric variable 
$\overline{\beta}{}^a{}_b := \beta^a{}_b/\lambda$. 

The twelve amplitude coefficients $a_1, a_2, b_1, b_2, c_1, c_2, d_1, d_2, 
p_1, p_2, q_1, q_2$ are not arbitrary but are related among themselves via the 
jump conditions for the electromagnetic field strength and the excitations 
at the boundaries $S_1$ and $S_2$. There are, as usual, twelve jump conditions 
-- six for every boundary surface. They read \cite{HO02}: 
\begin{eqnarray}
\left({\cal D}_{(1)} - {\cal D}_{(3)}\right)\,\vline\,
{\hbox{\raisebox{-1.5ex}{\scriptsize{$S_1$}}}}\wedge \nu = 0,\qquad
\tau_{A}\rfloor\left({\cal H}_{(1)} - {\cal H}_{(3)}\right)\,\vline\,
{\hbox{\raisebox{-1.5ex}{\scriptsize{$S_1$}}}}= 0,\label{jump1}\\
\left({B}_{(1)} - {B}_{(3)}\right)\,\vline\,
{\hbox{\raisebox{-1.5ex}{\scriptsize{$S_1$}}}}\wedge \nu  = 0,\qquad
\tau_{A}\rfloor\left({E}_{(1)} - {E}_{(3)}\right)\,\vline\,
{\hbox{\raisebox{-1.5ex}{\scriptsize{$S_1$}}}} = 0.\label{jump2}\\
\left({\cal D}_{(3)} - {\cal D}_{(2)}\right)\,\vline\,
{\hbox{\raisebox{-1.5ex}{\scriptsize{$S_2$}}}}\wedge \nu = 0,\qquad
\tau_{A}\rfloor\left({\cal H}_{(3)} - {\cal H}_{(2)}\right)\,\vline\,
{\hbox{\raisebox{-1.5ex}{\scriptsize{$S_2$}}}}= 0,\label{jump3}\\
\left({B}_{(3)} - {B}_{(2)}\right)\,\vline\,
{\hbox{\raisebox{-1.5ex}{\scriptsize{$S_2$}}}}\wedge \nu  = 0,\qquad
\tau_{A}\rfloor\left({E}_{(3)} - {E}_{(2)}\right)\,\vline\,
{\hbox{\raisebox{-1.5ex}{\scriptsize{$S_2$}}}} = 0.\label{jump4}
\end{eqnarray}
Here $\nu = dx$ is the 1-form density normal to the surfaces and $\tau_1 = 
\partial_y, \tau_2 = \partial_z$ ($A =1,2$) are the two vectors tangential 
to the boundaries. 

Substituting (\ref{E1})-(\ref{H3}), we find that some of the jump 
conditions are trivially satisfied since $B\wedge\nu = 0$ and ${\cal D}
\wedge\nu = 0$ in all the three regions. Accordingly, we are left
with only eight conditions which result from the continuity of $\tau_A
\rfloor{\cal H}$ and $\tau_A\rfloor E$ at the two boundaries. 
After some algebra, noting in particular that
\begin{eqnarray}
\varepsilon\varepsilon_0 \pm {\frac {k_1^\pm}\omega}\beta^2{}_3 =
\alpha_1\sqrt{\frac {\varepsilon_0}{\mu_0}}\,{\frac {k_1^\pm}{\omega}},
&\qquad&  \varepsilon\varepsilon_0 \mp {\frac {k_2^\pm}\omega}\beta^3{}_2 =
\alpha_2\sqrt{\frac {\varepsilon_0}{\mu_0}}\,{\frac {k_2^\pm}{\omega}},\\
{\frac {k_1^\pm}{\mu\mu_0\omega}} \mp \beta^2{}_3 = \alpha_1\sqrt{\frac 
{\varepsilon_0}{\mu_0}}, &\qquad& {\frac {k_2^\pm}{\mu\mu_0\omega}} \pm 
\beta^3{}_2 = \alpha_2\sqrt{\frac {\varepsilon_0}{\mu_0}},
\end{eqnarray}
with the abbreviations
\begin{equation}
\alpha_1 := \sqrt{{\frac \varepsilon\mu}\,[1 +(\overline{\beta}{}^2{}_3)^2]},
\qquad \alpha_2:= \sqrt{{\frac \varepsilon\mu}\,[1 +(\overline{\beta}{}^3{}_2)^2]},
\end{equation}
we can bring this system into the form of the {\it eight} equations:
\begin{eqnarray}
e^{-ik\ell}a_1 - e^{ik\ell}b_1 - \alpha_1 e^{-ik_1^+\ell}p_1 
+ \alpha_1 e^{ik_1^-\ell}q_1 &=& 0,\label{c1}\\
e^{-ik\ell}a_1 + e^{ik\ell}b_1 - e^{-ik_1^+\ell}p_1 
- e^{ik_1^-\ell}q_1 &=& 0,\label{c2}\\
e^{ik\ell}c_1 - e^{-ik\ell}d_1 - \alpha_1 e^{ik_1^+\ell}p_1 
+ \alpha_1 e^{-ik_1^-\ell}q_1 &=& 0,\label{c3}\\
e^{ik\ell}c_1 + e^{-ik\ell}d_1 - e^{ik_1^+\ell}p_1 
- e^{-ik_1^-\ell}q_1 &=& 0,\label{c4}\\
e^{-ik\ell}a_2 - e^{ik\ell}b_2 - \alpha_2 e^{-ik_2^+\ell}p_2 
+ \alpha_2 e^{ik_2^-\ell}q_2 &=& 0,\label{c5}\\
e^{-ik\ell}a_2 + e^{ik\ell}b_2 - e^{-ik_2^+\ell}p_2 
- e^{ik_2^-\ell}q_2 &=& 0,\label{c6}\\
e^{ik\ell}c_2 - e^{-ik\ell}d_2 - \alpha_2 e^{ik_2^+\ell}p_2 
+ \alpha_2 e^{-ik_2^-\ell}q_2 &=& 0,\label{c7}\\
e^{ik\ell}c_2 + e^{-ik\ell}d_2 - e^{ik_2^+\ell}p_2 
- e^{-ik_2^-\ell}q_2 &=& 0.\label{c8}
\end{eqnarray}
Thus, we can always choose 2 waves (inside or outside) the medium as independent
and find all the other waves in all three regions of space from the system
(\ref{c1})-(\ref{c8}). Obviously, the waves with a particular polarization are
only related to the waves of the same polarization. They do not mix with
the modes of a different polarization. 

The following three cases exhaust all possible situations: (i) we choose as 
primary the waves inside the medium, i.e., the amplitudes $p_A, q_A$ (with
$A=1,2$) are independent, and the amplitudes $a_A, b_A$ and $c_A, d_A$ outside
the matter are obtained from them as secondary, (ii) the waves in 
one vacuum region (for example, in the first one) are primary, then $a_A, b_A$ 
are independent and the waves in the matter $p_A, q_A$ and in the second vacuum
region $c_A, d_A$ are derived from them, (iii) the ingoing waves in the 
vacuum regions, $a_A, d_A$, are independent, then the waves in the matter 
$p_A, q_A$ and the outgoing waves $b_A, c_A$ are derived. The system 
(\ref{c1})-(\ref{c8}) can be straightforwardly solved for all these cases.

(i) Assuming $p_A, q_A$ to be the independent variables, see Table~\ref{table2},
we find from (\ref{c1})-(\ref{c8}) the amplitudes of the waves in the two vacuum 
regions:
\begin{eqnarray}
a_A &=& {\frac {e^{ik\ell}}2}\left[(1 + \alpha_A)e^{-ik_A^+\ell}p_A
+ (1 - \alpha_A)e^{ik_A^-\ell}q_A\right],\label{aA}\\
b_A &=& {\frac {e^{-ik\ell}}2}\left[(1 - \alpha_A)e^{-ik_A^+\ell}p_A
+ (1 + \alpha_A)e^{ik_A^-\ell}q_A\right],\label{bA}\\
c_A &=& {\frac {e^{-ik\ell}}2}\left[(1 + \alpha_A)e^{ik_A^+\ell}p_A
+ (1 - \alpha_A)e^{-ik_A^-\ell}q_A\right],\label{cA}\\
d_A &=& {\frac {e^{ik\ell}}2}\left[(1 - \alpha_A)e^{ik_A^+\ell}p_A
+ (1 + \alpha_A)e^{-ik_A^-\ell}q_A\right].\label{dA}
\end{eqnarray}
\begin{table}
\caption{\label{table2}Case 1. Given are the amplitudes in region 3, the rest
   is computed}
\begin{ruledtabular}
\begin{tabular}{|c|c|c|}\hline
  region 1&region 3& region 2\\ \hline\hline
  compute&$p_1,p_2$&compute\\
$\longrightarrow$&$\longrightarrow$&$\longrightarrow$\\ \hline
  compute&$q_1,q_2$&compute\\
$\longleftarrow$&$\longleftarrow$&$\longleftarrow$\\ \hline
\end{tabular}
\end{ruledtabular}
\end{table}
\begin{table}
\caption{\label{table3}Case 2. Given are the amplitudes in region 1. The rest
   is computed. There is also an equivalent case if only the amplitudes
   in region 2 are specified}
\begin{ruledtabular}
\begin{tabular}{|c|c|c|}\hline
  region 1&region 3& region 2\\ \hline\hline
$a_1,a_2$&compute&compute\\
$\longrightarrow$&$\longrightarrow$&$\longrightarrow$\\ \hline
$b_1,b_2$&compute &compute\\
$\longleftarrow$&$\longleftarrow$&$\longleftarrow$\\ \hline
\end{tabular}
\end{ruledtabular}
\end{table}
\begin{table}
\caption{\label{table4}Case 3: The amplitudes of the ingoing waves (with
   respect to region 3) are specified, the rest is computed}
\begin{ruledtabular}
\begin{tabular}{|c|c|c|}\hline
  region 1&region 3& region 2\\ \hline\hline
$a_1,a_2$&compute&compute\\
$\longrightarrow$&$\longrightarrow$&$\longrightarrow$\\ \hline
compute &compute&$d_1,d_2$\\
$\longleftarrow$&$\longleftarrow$&$\longleftarrow$\\ \hline
\end{tabular}
\end{ruledtabular}
\end{table}

(ii) Assume now that the waves in the left vacuum region are primary, see 
Table~\ref{table3}. Then $a_A, b_A$ are independent variables, and we find 
for the amplitudes in the medium and in the second vacuum region: 
\begin{eqnarray}
p_A &=& {\frac {e^{ik_A^+\ell}} {2\alpha_A}}\left[(1 + \alpha_A)
e^{-ik\ell}a_A - (1 - \alpha_A)e^{ik\ell}b_A\right],\label{p2A}\\
q_A &=& {\frac {e^{-ik_A^-\ell}} {2\alpha_A}}\left[- (1 - \alpha_A)
e^{-ik\ell}a_A + (1 + \alpha_A)e^{ik\ell}b_A\right],\label{q2A}\\
c_A &=& {\frac {e^{i(k_A^+ - k_A^-)\ell}} {2\alpha_A}}\left[K_Ae^{-2ik\ell}a_A
- i(1 - \alpha_A^2)\sin(k_A^+ + k_A^-)\ell\,b_A\right],\label{c2A}\\
d_A &=& {\frac {e^{i(k_A^+ - k_A^-)\ell}} {2\alpha_A}}\left[i(1 - \alpha_A^2)
\sin(k_A^+ + k_A^-)\ell\,e^{-2ik\ell}a_A + K_A^*b_A\right].\label{d2A}
\end{eqnarray}
Here we denoted 
\begin{eqnarray}
K_A &:=& 2\alpha_A\cos(k_A^+ + k_A^-)\ell + i(1 + \alpha_A^2)
\sin(k_A^+ + k_A^-)\ell,\label{KA}\\
\Delta_A &:=& K_AK_A^* = 4\alpha_A^2 + (1-\alpha_A^2)^2
\sin^2(k_A^+ + k_A^-)\ell.\label{DA}
\end{eqnarray}
The second quantity will be needed below. 
The star denotes complex conjugation as usual. 

(iii) Finally, if we assume, in accordance with Table~\ref{table2}, that the 
ingoing waves in the two vacuum regions, namely $a_A, d_A$, are independent, 
the amplitudes of the waves in matter $p_A, q_A$ and of the outgoing waves 
$b_A, c_A$ read: 
\begin{eqnarray}
p_A &=& {\frac {K_Ae^{-ik\ell}}{\Delta_A}}\left[(1 + \alpha_A)e^{-ik_A^+\ell}
a_A - (1 - \alpha_A)e^{ik_A^-\ell}d_A\right],\label{p3A}\\
q_A &=& {\frac {K_Ae^{-ik\ell}}{\Delta_A}}\left[(1 + \alpha_A)e^{-ik_A^+\ell}
a_A - (1 - \alpha_A)e^{ik_A^-\ell}d_A\right],\label{q3A}\\
b_A &=& {\frac {K_Ae^{-2ik\ell}}{\Delta_A}}\left[-i(1 - \alpha_A^2)\sin(k_A^+ 
+ k_A^-)\ell\,a_A + 2\alpha_Ae^{i(k_A^- - k_A^+)\ell}d_A\right],\label{b3A}\\
c_A &=& {\frac {K_Ae^{-2ik\ell}}{\Delta_A}}\left[2\alpha_Ae^{i(k_A^+ - k_A^-)\ell}
a_A - i(1 - \alpha_A^2)\sin(k_A^+ + k_A^-)\ell\,d_A \right].\label{c3A}
\end{eqnarray}

Both quantities (\ref{KA}) and (\ref{DA}) depend on the sums
\begin{equation}\label{sumk}
k_A^+ + k_A^- = {\frac {2n\omega}{c}}\,\sqrt{\overline{\beta}{}_A{}^2 + 1},
\end{equation}
with $\overline{\beta}{}_1 = \overline{\beta}{}^2{}_3, \overline{\beta}{}_2 
= -\overline{\beta}{}^3{}_2$. In practice, the magnetoelectric parameters are
rather small (typically of order of $10^{-4}-10^{-6}$), so with very high 
accuracy, we have $k_A^+ + k_A^- = {\frac {2n\omega}{c}} = 2k$. 

However, since the difference
\begin{equation}
k_A^+ - k_A^- = {\frac {2n\omega}{c}}\,\overline{\beta}{}_A\label{diffk}
\end{equation}
for the magnetoelectric matter is nontrivial, the amplitudes of the {\it 
left-moving waves in matter} are in general distinct from that of the 
{\it right-moving waves}. Similarly, the amplitudes of the outgoing waves in 
the two vacuum regions are different, in general. We have to check now if such 
a difference can be manifest in different field momentum densities and forces
in the two vacuum regions.

%%%%%%%%%%%%%%%%%%%%%%%%%%%%%%%%%%%%%%%%%%%%%%%%%%%%%%%%%%%
\section{Energy, momentum, and forces}\label{energydensity}
%%%%%%%%%%%%%%%%%%%%%%%%%%%%%%%%%%%%%%%%%%%%%%%%%%%%%%%%%%%

In vacuum, the energy-momentum 3-form of the electromagnetic field reads
\cite{HO02}
\begin{equation}
\Sigma_\alpha = {\frac 12}\left[F\wedge (e_\alpha\rfloor H) 
- (e_\alpha\rfloor F)\wedge H\right].
\end{equation}
Using the $(1+3)$-decomposition, with $F = E\wedge dt + B$ and $H= -{\cal H}
\wedge dt + {\cal D}$, we find explicitly for the temporal and spatial parts:
\begin{eqnarray}
\Sigma_{0} &=& u - dt\wedge s,\\
\Sigma_a &=& - p_a - dt\wedge S_a. 
\end{eqnarray} 
As we see, the energy-momentum generically decomposes into the four pieces: 
The {\it energy} density 3-form
\begin{equation}
  u:={\frac 1 2}\left(E\wedge{\cal D} + B\wedge{\cal
      H}\right)\,,\label{maxener}
\end{equation}
the {\it energy flux} density (or Poynting) 2-form
\begin{equation}
s:= E\wedge{\cal H}\,,\label{poynting}
\end{equation}
the {\it momentum} density 3-form
\begin{equation}
p_a:= -\,B\wedge(e_a\rfloor{\cal D})\,,\label{maxmom}
\end{equation}
and the Maxwell {\it stress} (or momentum flux density) 2-form of
the electromagnetic field
\begin{eqnarray}
S_a &:=&\frac{1}{{2}}\,\bigl[(e_{ a }\rfloor E)\wedge{\cal D} -(e_{a}
\rfloor{\cal D})\wedge E \nonumber\\ &&\hspace{3pt} +\,(e_{a}\rfloor
{\cal H})\wedge B -(e_{a}\rfloor B)\wedge {\cal H}\bigr]\,.\label{maxstress}
\end{eqnarray}

We will evaluate these expressions for the plane wave configurations
discussed in the previous section. Only the mean averaged (over a time period) 
quantities have a direct physical meaning. Using (\ref{E1})-(\ref{H1}),
we then find for the averaged quantities in the {\it first region}:
\begin{eqnarray}
<\!u\!> &=& {\frac {\varepsilon_0} 2}\sum_{A=1}^2(|a_A|^2 + |b_A|^2)
\,dx\wedge dy\wedge dz,\label{uleft}\\ \label{sl}
<\!s\!> &=& {\frac 1 {2\mu_0c}}\,\sum_{A=1}^2(|a_A|^2 - |b_A|^2)\,dy\wedge dz,\\
<\!p_a\!> &=& {\frac {\varepsilon_0} {2c}}\sum_{A=1}^2\left(\begin{array}{c}
(|a_A|^2 - |b_A|^2)dx\wedge dy\wedge dz\\ 0\\ 0\end{array}\right),\label{Pleft}\\
<\!S_a\!> &=& -\,{\frac {\varepsilon_0} 2}\sum_{A=1}^2\left(\begin{array}{c}
(|a_A|^2 + |b_A|^2)\,dy\wedge dz\\ 0\\ 0\end{array}\right).\label{Sleft}
\end{eqnarray}
The form of corresponding quantities in the {\it second region} is the same,
with the amplitudes $a_A$ replaced with $c_A$ and $b_A$ replaced with $d_A$. 

Consequently, it remains to calculate the moduli of the amplitudes of the 
incoming and outgoing waves, and compare the resulting quantities in the first
and in the second regions. We begin with the case (i) which directly corresponds
to the computations of Feigel. We take the virtual waves in the medium as 
primary fields, assuming the equal amplitudes $p_A = q_A$ for the left- and
right-movers, and then find from (\ref{aA})-(\ref{dA}) the waves in the two
vacuum regions: 
\begin{equation}
|a_A|^2 = |b_A|^2 = |c_A|^2 = |d_A|^2 = {\frac {|p_A|^2}2}\left[1 + \alpha_A^2
+ (1 - \alpha_A^2)\cos(k_A^+ + k_A^-)\ell\right].
\end{equation}
Accordingly, we find that the total momentum and the total energy flux are
vanishing in both vacuum regions, see (\ref{sl}) and (\ref{Pleft}), whereas 
the energy density and the stress densities are the same, see (\ref{uleft}) 
and (\ref{Sleft}). Recalling that the force acting on the boundary 
can be calculated as an integral  ${\cal F}_x = - \int <\!S_x\!>$, we then
conclude that the resulting force acting on the medium is zero, since on the
left boundary the normal vector points in the negative $x$-direction, and 
on the right boundary in the positive one. 

The same conclusion is derived when we analyse the case (ii) with the waves
of the equal amplitude travelling in the first or in the second vacuum regions.
When we evaluate the combined effect of their contributions, we again find the
zero resulting force acting on the medium. 

Somewhat different is the case (iii) when the ingoing waves (the right-moving
$a_A$ in the first (left) region and the left-moving wave in the second region 
with the same amplitude $d_A = a_A$) are considered as primary. Then for the 
outgoing waves we find from (\ref{b3A})-(\ref{c3A})
\begin{eqnarray}
|b_A|^2 &=& |a_A|^2\,\left[1 + {\frac {4\alpha_A}{\Delta_A}}(1 - \alpha_A^2)
\,\sin(k_A^+ + k_A^-)\ell\,\sin(k_A^+ - k_A^-)\ell\right],\label{bb}\\
|c_A|^2 &=& |a_A|^2\,\left[1 - {\frac {4\alpha_A}{\Delta_A}}(1 - \alpha_A^2)
\,\sin(k_A^+ + k_A^-)\ell\,\sin(k_A^+ - k_A^-)\ell\right].\label{cc}
\end{eqnarray}

As we can see, the contributions of the {\it outgoing} waves to the field
momentum are clearly {\it different} in the two vacuum regions. The difference
reads explicitly
\begin{equation}
|b_A|^2 - |c_A|^2 = {\frac {8|a_A|^2\alpha_A (1 - \alpha_A^2)} {\Delta_A}}
\,\sin(k_A^+ + k_A^-)\ell\,\sin(k_A^+ - k_A^-)\ell.\label{b-c}
\end{equation}
When $\beta^a{}_b = 0$, in view of (\ref{diffk}) the ``bath" of virtual waves 
around the sample is in equilibrium since then (\ref{b-c}) vanishes. Under this
assumption, the total momentum of the waves in both vacuum regions is obviously 
equal to zero. However, for  magnetoelectric matter, the mentioned ``bath" is 
still balanced in the sense that the field momentum carried by the waves in the 
left vacuum region is the {\it same} as that of the waves in the right 
vacuum region. Namely, substituting (\ref{bb}) and (\ref{cc}) into (\ref{Pleft}), 
we find that the momentum densities of the electromagnetic field {\it in both 
regions} are equal
\begin{equation}
<\!p_x\!> = -\,\sum_{A=1}^2{\frac {2\varepsilon_0|a_A|^2\alpha_A} {c\,\Delta_A}}
\,(1 - \alpha_A^2)\,\sin(k_A^+ + k_A^-)\ell\,\sin(k_A^+ - k_A^-)\ell
\,dx\wedge dy\wedge dz.\label{Px}
\end{equation}
At the same time, the stress is different in the two regions. When we
compute the corresponding force ${\cal F}_x = -\,\int <\!S_x\!>$, acting on 
the boundary of the sample, the resulting expressions will read for the 
first (left) and for the second (right) surfaces, respectively:
\begin{eqnarray}
{\cal F}_x^{\rm left} \!\!&=&\!\! \sum_{A=1}^2\varepsilon_0|a_A|^2
{\cal A}\,\left[1 + {\frac {2\alpha_A}{\Delta_A}}(1 - \alpha_A^2)
\,\sin(k_A^+ + k_A^-)\ell\,\sin(k_A^+ - k_A^-)\ell\right],\\
{\cal F}_x^{\rm right} \!\!&=&\!\! -\,\sum_{A=1}^2\varepsilon_0|a_A|^2
{\cal A}\,\left[1 - {\frac {2\alpha_A}{\Delta_A}}(1 - \alpha_A^2)
\,\sin(k_A^+ + k_A^-)\ell\,\sin(k_A^+ - k_A^-)\ell\right].
\end{eqnarray}
Here ${\cal A}$ is the area of the boundary surface (we assume the left
and right surfaces to be equal). Thus, there will be a nontrivial resulting
force acting on the sample in the direction of the magnetoelectric vector:
\begin{equation}
{\cal F}_x^{\rm left}+{\cal F}_x^{\rm right} = \sum_{A=1}^2{\frac {4\varepsilon_0
|a_A|^2{\cal A}\alpha_A}{\Delta_A}}(1 - \alpha_A^2)\,\sin(k_A^+ + k_A^-)\ell
\,\sin(k_A^+ - k_A^-)\ell.\label{force}
\end{equation}

At first sight, the results obtained, namely (\ref{Px}) and (\ref{force}),
provide a theoretical {\it support} for the possible {\it Feigel effect}.
For completeness, however, it is necessary to analyse also the situation 
when the directions of all waves are reversed, i.e., instead of assuming 
equal incoming waves, we should also consider the case of equal outgoing
waves. Fortunately, it is not necessary to perform a new computation. All
we need is to put $k\rightarrow -k$ in (\ref{E1})-(\ref{H3}) and then 
note that in the jump equations (\ref{c1})-(\ref{c8}) we have to change 
the sign of $\alpha_A\rightarrow -\alpha_A$. Then repeating the computations
of the amplitudes $p_A, q_A, b_A, c_A$, we arrive again to the solution 
(\ref{p3A})-(\ref{c3A}) with the replacements $k\rightarrow -k$ and 
$\alpha_A\rightarrow -\alpha_A$. As a result, the total momentum density
turns out to be again (\ref{Px}). [It is important to note that here we 
{\it do not} have to replace $\alpha_A\rightarrow -\alpha_A$, since $(|a_A|^2 
- |b_A|^2)$ is changed to $(|b_A|^2 - |a_A|^2)$ in (\ref{Pleft}), and analogously, 
$(|c_A|^2 - |d_A|^2)$ is changed to $(|d_A|^2 - |c_A|^2)$ in the similar formula
in the right region]. However, the resulting force computed from the stress on 
the left and right boundary surfaces will have the {\it opposite sign} (and 
equal magnitude) to that of (\ref{force}). Correspondingly, when we consider
both contributions together, the total force will be found to be equal 
to zero. In other words, in contrast to Feigel's result, the {\it 
magnetoelectric body will not move}, despite the presence of a certain 
asymmetry between the left- and right-moving waves in the matter. 

Several remarks are in order. The above conclusions are based on the evident 
symmetry that characterizes the ``bath" of the virtual photons (``vacuum 
fluctuations") in the regions 1 and 2: for each left-moving virtual photon 
there is an equal right-moving virtual photon. 

Furthermore, our observations demonstrate the difference between the two
irreducible parts of the magnetoelectric matrix $\beta^a{}_b$. Namely, assuming 
that only first (symmetric) irreducible part is nontrivial, we have $\beta^2{}_3
= \beta^3{}_2$, and then the waves of the different polarization contribute
to the above formulas with terms of the opposite sign since then
$\sin(k_1^+ - k_1^-)\ell = -\sin(k_2^+ - k_2^-)\ell$. The natural assumption
that the virtual waves of any polarization are produced with equal 
probability then leads to the conclusion that the net result will be zero
due to the mutual compensation of the contributions of the waves of different
polarizations. In other words, we find that the first irreducible part of
the magnetoelectric matrix is irrelevant for the possible Feigel effect. 
However, for the second (antisymmetric) part the situation is different. 
Then $\beta^2{}_3 = - \beta^3{}_2$, and the waves of both polarizations now
contribute with terms of the same sign. This was also observed in the 
original computation \cite{feigel} with the result obtained there which is
proportional to the skew-symmetric combination $\beta^2{}_3 - \beta^3{}_2$. 
This shows that the only the second (antisymmetric or pseudotrace) irreducible 
part of the magnetoelectric matrix (i.e., the 1-form $\check{\beta}$) is 
responsible for the possible Feigel effect. 

Finally, we can make some {\it predictions} going beyond the scheme of the 
virtual waves of the original Feigel picture. Let us assume that {\it real\/} 
electromagnetic waves are directed on the magnetoelectric sample from the
two sides. This can be easily achieved if we split the initial beam into
the two beams which are then, with a simple system of mirrors, directed
on the two opposite sides of the magnetoelectric sample. Then we are 
exactly in the situation describe as the case (iii) above. The outcome  
which we derived in this case is that the sample will be affected by a
nontrivial force (\ref{force}) from the electromagnetic fields falling on it. 
The sign and the magnitude of this force is determined by the polarization
of the infalling waves and by the magnetoelectric parameters $\beta^2{}_3, 
\beta^3{}_2$ of the medium. This effect can be directly verified 
experimentally. Moreover, one can use this real Feigel effect for the 
measurement of the magnetoelectric parameters $\beta^2{}_3, \beta^3{}_2$.
This method is drastically different and much simpler in its practical 
realization than the usual measurements of the magnetoelectric parameters
from the observation of magnetization of a medium in an external electric
field (resp., electric polarization of a medium in an external magnetic field). 

%%%%%%%%%%%%%%%%%%%%%%%%%%%%%%%%%%%%%%%%%%%%%%%%%%%%%%%%%%%
\section{Discussion and conclusion}\label{Concl}
%%%%%%%%%%%%%%%%%%%%%%%%%%%%%%%%%%%%%%%%%%%%%%%%%%%%%%%%%%%

In this paper we analysed the propagation of electromagnetic waves in a 
magnetoelectric medium. Our aim was to use this analysis for the discussion of 
the possibility of the Feigel effect \cite{feigel} that was predicted recently.

We derived a natural decomposition of the magnetoelectric matrix $\beta^a{}_b$
into the two irreducible parts: the first (tracefree symmetric), and the second 
(antisymmetric, or pseudotrace). The magnetoelectric matrix is tracefree. However,
the axion also describes the magnetoelectric effect, see (\ref{explicit'}) 
and (\ref{explicit2}). Being a 4D pseudoscalar, the axion can be formally treated
as a trace of generalized magnetoelectric $3\times 3$ matrix. We discussed the
physical properties of the axion piece elsewhere \cite{ax1,ax2} and here we 
assume (like in \cite{feigel} and in the previous work \cite{emtme}) that the 
axion piece is absent. Ultimately we find that only the second part (i.e., the 
magnetoelectric pseudotrace or antisymmetric part) can be responsible for
the possible Feigel effect. Physically this result appears to be quite 
natural because the magnetoelectric pseudotrace 1-form $\check{\beta}$
selects a distinguished direction in space. Along this direction, we then 
subsequently find the nontrivial forces and electromagnetic momenta
acting on the magnetoelectric medium. 

We show that taking into account the finite size of a magnetoelectric sample,
one can reduce the problem of computation of the electromagnetic energy, the
momentum and the forces to the vacuum regions just outside the sample. We then
demonstrate that the net force on the sample turns out to be zero for the 
careful count of the contributions from all the the virtual electromagnetic
waves excited in the medium. 

At the same time, we discover that for {\it real} electromagnetic waves, that
fall along the direction of the pseudotrace $\check{\beta}$ on a sample from 
the two opposite sides, is nontrivial. Accordingly, it would make the
magnetoelectric sample move along $\check{\beta}$, indeed, as was predicted 
in the original work \cite{feigel}. We propose to use this observation for 
the measurement of the magnetoelectric susceptibilities. Such an method could 
provide a useful alternative scheme for the measurements of the magnetoelectric 
parameters, along with the traditional methods based on the measurements of 
polarization induced in an external magnetic field (or of magnetization 
induced by an external electric field). 

In the current paper, we have confined our attention (like also Feigel in
\cite{feigel}) to the waves induced along $\check{\beta}$. Strictly speaking,
the complete analysis requires also to take into account the waves that move
along other directions. The extension of the above results to such waves is
straightforward, although the formulas (especially those that relate the 
waves in the media and in vacuum at the boundary surfaces) turn out to be 
appreciable longer. The corresponding analysis, however, shows that the 
main contribution to the possible Feigel effect comes from the waves moving
along $\check{\beta}$, so the additional modes do not change the conclusions
qualitatively. 

{\bf Acknowledgment}. This work was supported by the Deutsche 
Forschungsgemeinschaft (Bonn), project He~528/21-1.

\end{document}